\documentstyle[emulateapj,epsfig]{article} 
\begin{document}

% Definitions 
\newcommand{\msol}{M{$_{\odot}$}} 
\newcommand{\lsol}{L{$_{\odot}$}}
\newcommand{\rsol}{R{$_{\odot}$}}
\newcommand{\kms}{km~s{$^{-1}$}} 
\newcommand{\cmq}{cm{$^{-2}$}} 
\newcommand{\cmc}{cm{$^{-3}$}} 
\newcommand{\hii}{H$_2$}
\newcommand{\hcop}{HCO$^+$($1\rightarrow 0$)}
\newcommand{\hicop}{H$^{13}$CO$^+$($1\rightarrow 0$)}
\newcommand{\hicn}{H$^{13}$CN($1\rightarrow 0$)}
\newcommand{\sio}{SiO(v=0, $2\rightarrow 1$)}
\newcommand{\ho}{H$_2$O}
\newcommand{\degr}{$^{\circ}$}
\newcommand{\asec}{$^{\prime \prime}$}                   %arcsec
\newcommand{\um}{$\mu$m}                                 %micron
\newcommand{\mdot}{\.{M}}
\newcommand{\msunyr}{$M_{\odot}$ yrs$^{-1}$}
\newcommand{\lsim}{\;\lower.6ex\hbox{$\sim$}\kern-7.75pt\raise.65ex\hbox{$<$}\;}
\newcommand{\gsim}{\;\lower.6ex\hbox{$\sim$}\kern-7.75pt\raise.65ex\hbox{$>$}\;}

	% End definitions
\lefthead{Molinari et al.}
\righthead{Detection of the 62 \um\ \ho\ ice feature toward HH\,7}

\title{Detection of the 62 \um\ crystalline \ho\ ice feature in 
emission toward HH\,7 with ISO-LWS\footnotemark[0]}

\author{Sergio Molinari}
\affil{Infrared Processing and Analysis Center, California 
Institute of Technology, MS 100-22, Pasadena, CA 91125, USA}
\author{Cecilia Ceccarelli}
\affil{Laboratoire d'Astrophysique, Observatoire de Grenoble - BP 53, 
F-38041 Grenoble cedex 09, France}
\author{Glenn J. White}
\affil{Department of Physics, Queen Mary and Westfield 
College, University of London, Mile End Road, London E1 4NS, UK}
\affil{Stockholm Observatory, S-133 36 - Saltsj\"osbaden, Sweden}
\author{Paolo Saraceno}
\affil{CNR-Istituto di Fisica dello Spazio Interplanetario, Area di 
Ricerca Tor Vergata, via Fosso del Cavaliere I-00133 Roma, Italy}
\author{Brunella Nisini and Teresa Giannini}
\affil{Osservatorio Astronomico di Roma, via Frascati 33, I-00044 
Monte Porzio, Italy}
\author{Emmanuel Caux}
\affil{CESR CNRS-UPS, BP 4346, F-31028 Toulouse Cedex 04, France}

%\author{Sergio Molinari \altaffilmark{1}}
%\author{Cecilia Ceccarelli \altaffilmark{2}}
%\author{Glenn J. White \altaffilmark{3,4}}
%\author{Paolo Saraceno \altaffilmark{2}}
%\author{Brunella Nisini \altaffilmark{6}}
%\author{Teresa Giannini \altaffilmark{6}}
%\author{Emmanuel Caux} \altaffilmark{7}
%\altaffiltext{1}{Infrared Processing and Analysis Center, California 
%Institute of Technology, MS 100-22, Pasadena, CA 91125, USA}
%\altaffiltext{2}{Laboratoire d'Astrophysique, Observatoire de Grenoble - 
%BP 53, F-38041 Grenoble cedex 09, France}
%\altaffiltext{3}{Department of Physics, Queen Mary and Westfield 
%College, University of London, Mile End Road, London E1 4NS, UK}
%\altaffiltext{4}{Stockholm Observatory, S-133 36 - Saltsj\"osbaden, 
%Sweden}
%\altaffiltext{5}{CNR-Istituto di Fisica dello Spazio Interplanetario, 
%Area di Ricerca Tor Vergata, via Fosso del Cavaliere I-00133 Roma, 
%Italy}
%\altaffiltext{6}{Osservatorio Astronomico di Roma, via Frascati 33, 
%I-00044 Monte Porzio, Italy}
%\altaffiltext{7}{CESR CNRS-UPS, BP 4346, F-31028 Toulouse Cedex 04, 
%France}

\footnotetext[0]{Based on observations with ISO, an ESA project
with instruments funded by ESA Member States (especially the PI
countries: France, Germany, the Netherlands and the United Kingdom) with
the participation of ISAS and NASA.}

\begin{abstract}
We report the detection of the 62 \um\ feature of {\it crystalline}
water ice in emission towards the bow-shaped Herbig-Haro object HH\,7.
Significant amounts of far infrared continuum emission are also detected
between 10 and 200 \um, so that Herbig-Haro objects cease to be pure
emission-line objects at FIR wavelengths.  The formation of crystalline
water ice mantles requires grain temperatures T$_{gr}\gsim$ 100 K at the
time of mantle formation, suggesting that we are seeing material
processed by the HH\,7 shock front. The deduced ice mass is $\sim~
2\times 10^{-5}$ \msol\ corresponding to a water column density
N(\ho)$\sim~ 10^{18}$ \cmq; an estimate of the [\ho]/[H] abundance
yields values close to the interstellar gas-phase oxygen abundance. The
relatively high dust temperature and the copious amounts of gas-phase
water needed to produce the observed quantity of crystalline water ice,
suggest a scenario where both dissociative and non-dissociative shocks
co-exist. The timescale for ice mantle formation is of the order of
$\sim 400$ years, so that the importance of gas-phase water cooling as a
shock diagnostic may be greatly diminished.

\end{abstract}

\keywords{(ISM:) dust, extinction --- ISM: Herbig-Haro objects --- 
ISM: individual (HH\,7) --- ISM: lines and bands --- infrared: ISM: 
continuum --- infrared: ISM: lines and bands}

\setcounter{footnote}{0}
\section{Introduction}
\label{introduction}
Herbig-Haro objects (HH; Haro~\cite{H50}, Herbig~\cite{H51}) are
emission-line objects acting as signposts for the shock regions (Draine,
Roberge \& Dalgarno~\cite{DRD83}; Hollenbach \& McKee~\cite{HM89})
originating at the interface between stellar winds accelerated by Young
Stellar Objects (YSOs) and the circumstellar or cloud ambient material.
The temperature of dust grains in these shock regions can raise to only
a few hundred degrees at most in dissociative shocks (Hollenbach \&
McKee~\cite{HM89}), so that their thermal emission cannot be detected
below 10 \um. 

The instruments on board the {\it Infrared Space Observatory} satellite 
(ISO, Kessler et al.~\cite{Ketal96}) opened unprecedented possibilities
for  far infrared continuum studies of cold objects, including HH
objects.  In this {\it Letter} we present the data obtained with the
Long (LWS, Clegg et al.~\cite{Clegg96}) and Short (SWS, de Graauw et
al.~\cite{dGetal96}) Wavelength Spectrometer towards HH\,7, the leading
bow-shaped shock of the HH\,7-11 chain emanating from the YSO SVS\,13,
in the star forming region NGC\,1333 in Perseus (d=350 pc). Details
about the observations and data reduction are given elsewhere (Molinari
et al.~\cite{Metal99}). In Sect.~\ref{results} the additional data
analysis procedures we adopted to derive a reliable continuum spectrum
for HH\,7 are discussed. Nomenclature for the ten LWS detectors is
described in the ISO Data User Manual ({\it
http://www.iso.vilspa.esa.es/manuals/lws\_idum5}; detectors SW1 to SW5 
(43-90 \um), and LW1 to LW5 (80-197 \um), are sometimes referred to as
``short'' and ``long''  wavelength detectors respectively. 

\section{Results}
\label{results}

The LWS beam centered on HH\,7, also includes object HH\,8 somewhat
20\asec\ off-axis and HH\,10 at the edge of the beam; these objects are
however fainter than HH\,7 at 2 \um\ (Molinari et al.~\cite{Metal99})
and the beam profile suppresses their possible contribution even more.
Apart from a contamination by the strong nearby source SVS\,13, which
will be discussed in detail in Sect.~\ref{continuum}, it is plausible to
assume that HH\,7 dominates the observed spectrum.

\subsection{The 62 \um\ feature}
\label{feature62}

In Fig.~\ref{hh7_62umplot} the complete spectra  observed towards HH\,7
and SVS\,13 are shown. A broad feature extending from roughly 50 to 70
\um\ is clearly visible in the HH\,7 spectrum. For clarity the SW3
detector spectrum, with an offset applied to align it with the 
continuum of the adjacent detectors, is shown with plus symbols. The
primary concern was to make sure that the 50-70\um\ feature was not a
result of a residual instrumental effect and that it is intrinsic to 
HH\,7. The passband calibration is known to be somewhat  inaccurate for
detector SW1 (it may be worse than 50\% both in absolute and relative
terms) which is the most sensitive to transients. Detectors SW2, SW3 and
SW4 however, are stable in relative (passband) terms and a conservative
figure of 30\% can be assumed for their absolute calibration accuracy. 
The possibility that the broad feature visible on
Fig.~\ref{hh7_62umplot} may be due to the near-IR leaks present in the
LWS detectors filters ({\it
http://isowww.estec.esa.nl/notes/lws\_0197.html}) can also be excluded.
Finally, this feature might result from contamination effects from the
bright nearby source SVS\,13, which is the candidate exciting source for
HH\,7; these effects will be discussed in detail in the
Sect.~\ref{continuum}.

\smallskip
\vspace{7cm}
\includegraphics{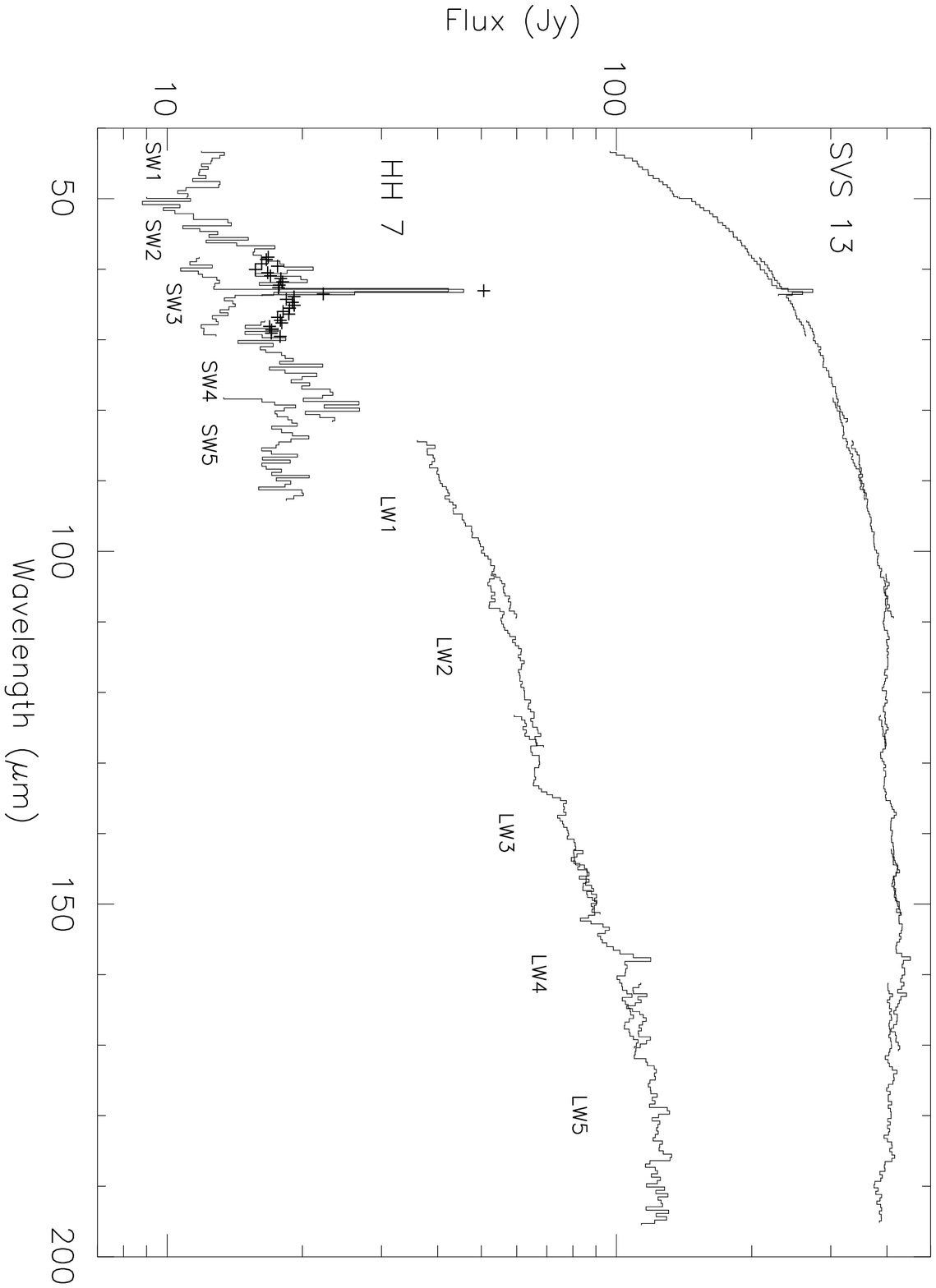}
{\small\noindent Fig. 1 --
%\caption
LWS averaged spectra observed towards HH\,7 (bottom) and SVS\,13 (top).
The different portions  are the 10 LWS detectors, labeled with the names
used in the text. The plus symbols for HH\,7 represent detector SW3 with an applied offset of 5 Jy to bring it in  line with the adjacent detectors.
}
\label{hh7_62umplot} 
%\end{figure}
\smallskip

Here we just note here that if the 50-70 \um\ feature seen on the HH\,7
spectrum was due to a fraction of the flux emitted by the nearby
contaminating source SVS\,13, we would expect to see the same feature at
a comparable ``line-to-continuum'' ratio on the SVS\,13 continuum
spectrum; Fig.~\ref{hh7_62umplot} clearly shows that this is not the
case. We conclude that the 50-70 \um\ feature is real and intrinsic to
HH\,7 and we consider it as a 2$\sigma$ detection, assuming that the
noise of the feature is equal to half the gap between detectors SW3 and
SW2. We identify this as the 62\um\ feature due to the longitudinal
acoustic modes of crystalline water ice (Bertie, Labbe \&
Whalley~\cite{BLW69}), observed for the first time by Omont et al.
(\cite{Oetal90}) in the expanding envelopes of post-AGB stars; ISO-LWS
has detected this feature in the spectra of similar objects
(Barlow~\cite{B98}) and of a few Herbig Ae/Be stars (Waters \&
Waelkens~\cite{WW98}, Malfait et al.~\cite{Malfait}). It is the first
time this feature has been detected towards Herbig-Haro objects.

\subsection{The 2-200 \um\ Continuum}
\label{continuum}

Several SWS line scans were used to estimate the continuum at different
wavelengths for $\lambda <$40 \um. The line-free portions of each
individual spectrum were used to build flux histograms, and a gaussian
was then fitted to the core of the distribution to obtain the centroid
(average flux) and the standard deviation; the latter was then divided
by the square root of the number of points in histogram to get an
estimate of the uncertainty. These uncertainties reflect the internal
accuracy of the estimates; infact the true uncertainty may be higher. In
particular, all data shortward of $\sim$ 10 \um\ are at the detection
limit of the SWS and they will be treated as 1$\sigma$ upper limits.

LWS scans were averaged using the ISO Spectral Analysis Package (ISAP,
{\it http://www.ipac.caltech.edu/iso/isap/isap.html}) using a median
clipping algorithm which rejected outlying data points  due to
incomplete removal of glitches from cosmic rays hits. However,
contamination effects in the HH\,7 spectrum due to the nearby ($\sim$
70\asec\ NW) candidate exciting source SVS\,13 must be taken into
account to assess if the observed continuum (Fig.~\ref{hh7_62umplot} is
all due to HH\,7; for this purpose we used an irregularly sampled raster
map of Mars taken as part of the LWS calibration programme. We first
used the LIA ({\it http://www.ipac.caltech.edu/iso/lia/lia.html})
routine INSPECT\_RASTER to  determine the relative position of HH\,7 and
SVS\,13 in the [Y,Z] spacecraft frame of reference. For each Mars raster
position along the direction connecting the two sources, we averaged the
Mars spectra detector by detector; each detector average at the various
off-axis positions was then  ratioed to the analogue detector average of
the on-axis spectra, and a set of ten contamination factors for each
position was obtained. A spline interpolation was then used to estimate
these factors (one per detector) at an off-axis distance of 68\asec, the
distance between HH\,7 and SVS\,13; we find values 0.024, 0.025, 0.037,
0.05, 0.075, 0.09, 0.10, 0.11, 0.13 and 0.15, which increase with
wavelength as expected due to diffraction. We multiplied the observed
SVS\,13 spectra by these numbers, and the resultant spectrum was
subtracted from the observed HH\,7 spectrum. 

This method for estimating the contamination from nearby sources 
inevitably suffers from the irregular sampling of the Mars raster map. 
Raster points did not lay exactly along the HH\,7-SVS\,13 direction on
the focal plane, and the interpolation between the correction factors
estimated at each relevant raster position introduced an additional
uncertainty. At the end of the procedure (see Fig.~\ref{modelfit}) we 
find worse alignment between adjacent LW detectors, and detector SW5 
goes to negative flux values (not reported on Fig.~\ref{modelfit}).  We
believe this may be due to the fact that as the magnitude of the 
applied contamination correction increases with wavelength, so it  does
the associated uncertainty; in particular, detector SW5 is in  the
critical region where the contamination correction starts to be  high
(2-3 times higher than for SW2 and SW4) while the observed signal  is
still low (similar to SW2), so that a slightly overestimate  of the
contamination fraction is enough to bring its corrected values  below
zero. The LWS LW detectors are those for which the contamination 
fractions are higher and for which diffraction and source's extension 
effects (which here have been neglected) are more severe. An additional 
complication, which does not affect the SW detectors, is that heavy 
fringing is observed (which has been here removed using standard tools 
available in ISAP); this has the effect of modulating the beam size as a
function of wavelength even within individual detector bands. We did not
take this into account, instead deriving a single contamination factor
per detector; multiplying or dividing by a constant, however,  has the
effect of changing the slope. Hence, although the  observed continuum
levels are high enough to ensure that an important part of the observed
signal can confidently be assigned to HH\,7, the exact absolute fluxes
and spectral shape in this  wavelength range remain highly uncertain. 

On the short wavelength side of the LWS spectrum on the other hand, 
the existence of intrinsic continuum emission from HH\,7 depends more 
critically on the particular value of the contamination factors; 
doubling this factors would lower the spectrum to negative flux 
levels. If, however, the $\lambda\leq$ 80 \um\ flux observed towards 
HH\,7 were due to contamination, we would expect to see the 62 \um\ 
emission feature also on the continuum spectrum of SVS\,13 and with 
a comparable ``line/continuum''; instead, no trace of such a feature 
is seen on the SVS\,13 continuum (see Fig.~\ref{hh7_62umplot}). We 
conclude that FIR emission which is intrinsic to HH\,7 has been 
detected, showing for the first time that Herbig-Haro objects cease 
to be exclusively emission-line objects at FIR wavelengths.

\section{Discussion}
\label{discussion}
In the following discussion we will assume that the FIR and mm continuum
arise from the same region of space, based on the spatial coincidence
between the HH objects and the mm emission distribution; currently
available 50-100 \um\ data on the region (Harvey et al.~\cite{Hetal84},
\cite{Hetal98}; Jennings et al.~\cite{Jetal87}) however, do not have
enough spatial resolution or sensitivity to support this claim. 

The spectral energy distribution (SED) from HH\,7 will be modeled as
thermal emission from dust grains composed of a silicate core and a
water ice mantle. Absorptivities between 2 and 300 \um\ were computed
with Mie theory in the formulation of Wickramasinghe (\cite{W67}),
using  the complex refractive indices for crystalline water ice (Bertie
et al.~\cite{BLW69}) and silicate (Draine~\cite{D85}). Longward of 300
\um\ the silicate absorptivities by Draine (\cite{D85}) were adopted;
inclusion of water ice mantles may steepen the slope of the Q$_{abs}$ vs
$\lambda$ relationship (Aannestad~\cite{A75}), possibly leading to
underestimation of the dust mass. Radiative transfer is approximated
with an analytical treatment where dust is distributed on a sphere which
is  characterised by radial density and temperature gradients
(Noriega-Crespo, Garnavich \& Molinari~\cite{NCGM98}). The presence of a
cold dust clump centered on the location of HH\,7 and extending over a
larger area than the one traced by the optical or near-IR emission
(Lefloch et al.~\cite{Letal98}), justifies a treatment where a dust
clump is centrally heated by the HH\,7 shock. Fig.~\ref{modelfit}
presents the complete SED  towards HH\,7, after removal of the
contamination from SVS\,13 (see  Sect.~\ref{continuum}). The LWS
detectors SW1 to SW4 have been rescaled  to a common level preserving
their original mean value; no other  rescaling was made to the rest of
the LWS spectra. The continuum points  from the SWS spectra and an
estimate for the 1.25 mm flux integrated on the model fit area (see
below) are also reported; the latter point has to be considered an upper
limit since it is contaminated by SVS\,13 just in the same way it
happens for the LWS observations. We find that it is impossible to 
globally fit the observed SED; considering the  discussion in
Sect.~\ref{continuum}, we then  decided to give higher priority  to the
short wavelength LWS detectors spectra also because these are the  ones
showing the 62 \um\ feature, and more closely match each other (not 
considering the applied rescaling).

While the overall appearance of the continuum depends on the radius of
the dust clump, its mass and temperature gradient, the
``line/continuum'' ratio of the 62 \um\ feature depends on the relative
size of the ice mantle with respect to the grain core. The 62 \um\ water
ice feature is well fitted by adopting a core radius of 0.07 \um\ and a
total core + mantle radius of 0.1 \um. The model also suggests that the
short wavelength end of the LWS spectrum can be identified with one side
of the 45 \um\ water ice feature (see also Dartois et
al.~\cite{Detal98}), also in emission; although detector SW1 is the one
that experiences the most severe problems in calibration-related
aspects, the further agreement of the model with the $\lambda\sim$38.4
\um\ SWS point makes this a plausible possibility. The importance of the
detection of the 62 \um\ feature is that it is specific to crystalline
water ice, as opposed to the 45 \um\ feature which is instead predicted
for both crystalline and amorphous ice (e.g. Moore et
al.~\cite{Metal94}). Most of the  laboratory studies (e.g., Smith et
al.~\cite{Setal94} and references  therein) show that ice deposited onto
grains at T$\lsim$100 K is in  amorphous state (but see Moore et al.
\cite{Metal94} who instead obtain  crystalline ice in these conditions)
also for deposition rates as high  as 20 \um\ h$^{-1}$ (L\'eger et
al.~\cite{Letal83}); the 62\um\ feature  is then an important indicator
of the thermal history of the grains  because it appears at T$\gsim$100
K when an irreversible transition  from amorphous to crystalline state
commences. 

\smallskip
\vspace{7cm}
\includegraphics{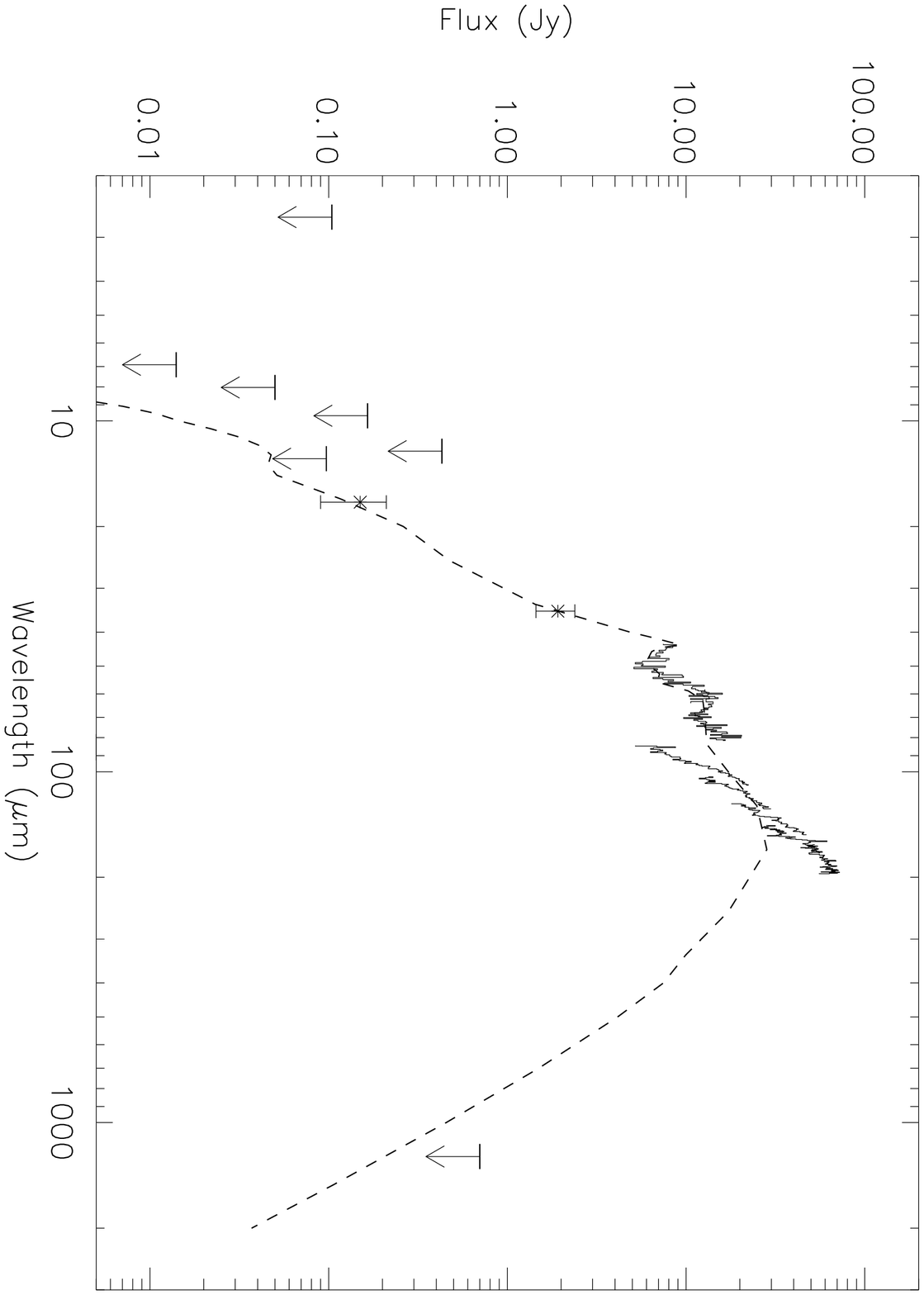}
{\small\noindent Fig. 2 --
%\caption
Complete spectral energy
distribution towards HH\,7. The full lines represent the LWS spectra
after applying the contamination corrections discussed in the text.
Detectors SW1 to SW4 have been rescaled to a common level preserving
their original mean value.}
\label{modelfit} 
%\end{figure}
\smallskip

The fit plotted in Fig.~\ref{modelfit} is obtained adopting a dust clump
radius of 0.06 pc, which approximates the radius of the average LWS beam
size at a distance of 350 pc, and the size of the 1.25 mm continuum
emission area (Lefloch et al.~\cite{Letal98}). The density is assumed
constant and equal to 6$\times 10^{-4}$ \cmc\ while the temperature
varies from $\sim$10 to 200 K with a $\sim-$0.4 power-law radial
gradient. The bolometric luminosity obtained integrating the model fit
is L$_{fit}$= 3.7 \lsol, while integration of the
contamination-corrected SED (Sect.~\ref{continuum}) yields L$_{sed}\lsim
4.5$ \lsol\ (since $\lambda<10$ \um\ and $\lambda$= 1.25 mm data are to
be considered upper limits); this is about a factor 30 higher than the
cooling via atomic and molecular lines observed towards HH\,7 (Molinari
et al.~\cite{Metal99}). The fitted model predicts that the bulk of the
62 \um\ feature is emitted by dust at temperatures T$\gsim$30 K
concentrated inside a 4\asec-radius region centered on HH\,7, a size
comparable to that of the optical and near-IR emission which traces the
shock front. Since these grains should have experienced a rise in
temperature to values $\gsim$100 K in order for ice mantles to be in
crystalline state, it is plausible that {\bf the 62 \um\ feature
originates from dust which has been processed by the HH\,7 shock}.  The
dust mass in the 4\asec-radius region centered on HH\,7 amounts to
5$\times 10^{-5}$ \msol\ (the total dust mass implied by the model fit
is $\sim$ 0.035 \msol); the relative proportion of core and mantle (the
core has 70\% of the total grain radius) implies a water ice mass of
$2\times 10^{-5}$ \msol, or a \ho\ column density $\sim~ 1.1\times
10^{18}$ \cmq. \hii\ pure rotational lines (Molinari et
al.~\cite{Metal99}) suggest N(\hii) $\sim~ 4.4\times 10^{20}$ \cmq\ in
the same 4\asec-radius region, implying [\ho]/[H]$\sim~
1.25\times10^{-3}$, or a factor $\sim$ 4 higher than the interstellar O
gas-phase abundance (Meyer et al.~\cite{MJC98}). However, this number
should be regarded as an upper limit because our model assumes that {\it
all} dust grains are coated  with ice mantles, which is not necessarily
true; besides, we cannot  exclude the presence of cold \hii\ (T$\lsim$
100 K) which our ISO  observations (Molinari et al.~\cite{Metal99})
would not trace. This water abundance, even if considered only as an
order-of-magnitude estimate, is however much higher than the gas phase
water abundance ([\ho]/[H]$\lsim~ 10^{-5}$) deduced from FIR lines
(Molinari et al.~\cite{Metal99}), and it would essentially imply that
most of the oxygen is locked into water ice. 

Due to the uncertainty about the 45 \um\ feature (see above), it hard to
tell from the observational viewpoint whether gas-phase water produced
behind the HH\,7 shock front (Kaufman \& Neufeld~\cite{KN96}) is
deposited onto bare warm grains, or pre-existing ice mantles are
warmed-up during the passage of a relatively gentle shock front. The ice
optical constants that we used (Bertie et al.~\cite{BLW69}) are from
laboratory samples obtained by direct deposition at T=173 K and
subsequent cooling to 100 K; hence the good simultaneous fit of the 62
and 45\um\ features would tend to support the first scenario. The
physical condition behind low velocity (v$_s\lsim$40 \kms),
non-dissociative, shocks are favourable for the rapid gas-phase
incorporation of atomic oxygen into water (Draine, Roberge \&
Dalgarno~\cite{DRD83}, Kaufman \& Neufeld~\cite{KN96}).  Subsequent
freezing onto grains (Bergin, Neufeld \& Melnick~\cite{BNM98},
\cite{BNM99}) in the cooling post-shock region could produce the
observed water ice mantles, also explaining the minor role played by
gas-phase water in the cooling of the HH\,7 shock (Molinari et
al.~\cite{Metal99}). However, non-dissociative shocks are unable to
raise the grain temperature to the T$\gsim$100 K (Draine, Roberge \&
Dalgarno~\cite{DRD83}) needed to explain the presence of  crystalline
ice mantles; a dissociative component, whose generated intense UV field 
(Hollenbach \& McKee~\cite{HM89}) is much more efficient in heating the
grains, is needed.  The co-existence of both types of shock is indeed an
expected feature of bow shocks like HH\,7 (Smith \& Brand~\cite{SB90})
and it is independently supported by FIR lines studies (Molinari et
al.~\cite{Metal99}). However, we point out that this relies on the
general result that crystalline ice requires relatively high
temperatures to form; if  crystalline ice can also form at low (T$<$ 20
K) temperature (Moore et  al.~\cite{Metal94}) then a dissociative shock
component may not be  needed.

The alternative possibility that pre-existing amorphous water ice
mantles are heated up during the passage of the shock front seems
unlikely since the mantles are easily destroyed by grain-grain
collisions once the shock velocities exceed $\sim$ 15 \kms\ (Caselli,
Hartquist \& Havnes~\cite{CHH97}).

We can derive an order-of-magnitude estimate for the timescale
$\tau_{ice}$ of the formation of water ice mantles, dividing the linear
size of the post-shock region  by the shock velocity; assuming a maximum
shock velocity of $\sim$ 40 \kms\ for the HH\,7 shock (Solf \&
B\"ohm~\cite{SB87}, Molinari et al.~\cite{Metal99}) and a linear extent
of $\sim$ 10\asec\ for the HH\,7  post-shock region as estimated from
H$_2$ 2.12\um\ images (Garden et al.~\cite{Getal90}, Everett \cite{E97},
Molinari et al~\cite{Metal99}), we estimate $\tau_{ice}\sim$ 400 yrs, a
factor $\sim$ 250 less compared to theoretical predictions for ice
mantles formation behind non-dissociative shocks (Bergin, Neufeld \&
Melnick~\cite{BNM99}). This very short timescale for ice mantle
formation would make the gas-phase water cooling practically irrelevant
as a shock diagnostic. 

\noindent
{\bf Acknowledgements:}
We thank B. Swinyard for providing us with the calibrated data for the
Mars raster map we used to estimate the contamination corrections, and
B. Lefloch for providing us with the integrated flux from his  published
1.25 mm map of NGC 1333. We also thank A. Noriega-Crespo and E. Sturm
for sharing their insight in SWS data. The ISO Spectral Analysis Package
(ISAP) is a joint development by the LWS and SWS Instrument Teams and
Data Centers. Contributing institutes are CESR, IAS, IPAC, MPE, RAL and
SRON.


\begin{thebibliography}{}

\bibitem[1975]{A75}
   Aannestad, P.A. 1975, \apj, 200, 30
\bibitem[1998]{B98}
   Barlow, M.J. 1998, \apss, 255, 315
\bibitem[1998]{BNM98}
   Bergin, E.A., Neufeld, D.A., Melnick, G.J. 1998, \apj, 499, 777
\bibitem[1999]{BNM99}
   Bergin, E.A., Neufeld, D.A., Melnick, G.J. 1999, \apjl, 510, 145
\bibitem[1969]{BLW69}
   Bertie, J.E., Labbe, H.J., Whalley, E. 1969, J. Phys. Chem., 50, 4501
%\bibitem[1993]{BNS93}
%   B\"ohm, K.-H., Noriega-Crespo, A., Solf, J. 1993, \apj, 416, 647
\bibitem[1997]{CHH97}
   Caselli, P., Hartquist, T.W., Havnes, O. 1997, \aap, 322, 296
\bibitem[1996]{Clegg96}
   Clegg, P.E., Ade, P.A.R., Armand, C., et al. 1996, \aap, 315, L38
\bibitem[1998]{Detal98}
   Dartois, E., Cox, P., Roelfsema, P.R., et al. 1998, \aap, 338, L21
\bibitem[1996]{dGetal96}
   de Graauw, T., Haser, L.N., Beintema, D.A., et al. 1996, \aap, 315, L49
\bibitem[1985]{D85}
   Draine, B.T. 1985, \apjs, 57, 587
\bibitem[1983]{DRD83}
   Draine, B.T., Roberge, W.G., Dalgarno, A. 1983, \apj, 264, 485
\bibitem[1997]{E97}
   Everett, M.E. 1997, \apj, 478, 246
\bibitem[1990]{Getal90}
   Garden, R.P., Russel, A.P.G., Burton, M.G. 1990, \apj, 354, 232
%\bibitem[1996]{G96}
%   Gredel, R. 1996, \aap, 305, 582
\bibitem[1950]{H50}
   Haro, G. 1950, \aj, 55, 72
\bibitem[1984]{Hetal84}
   Harvey, P.M., Wilking, B.A., Joy, M. 1984, \apj, 278, 156
\bibitem[1998]{Hetal98}
   Harvey, P.M., Smith, B.J., Di Francesco, J., Colom\`e, C. 1998, \apj, 499, 294
\bibitem[1951]{H51}
   Herbig, G.H. 1951, \apj, 113, 697
%\bibitem[1979]{HM79}
%   Hollenbach, D., McKee, C.F. 1979, \apjs, 41, 555
\bibitem[1989]{HM89}
   Hollenbach, D., McKee, C.F. 1989, \apj, 342, 306
\bibitem[1987]{Jetal87}
   Jennings, R.E., Cameron, D.H.M., Cudlip, W., Hirst, C.J. 1987, \mnras, 226, 461
\bibitem[1996]{KN96}
   Kaufman, M.J., Neufeld, D.A. 1996, \apj, 456, 611
\bibitem[1996]{Ketal96}
   Kessler, M.F., Steinz, J.A., Anderegg, M.E., et al. 1996, \aap, 315, L27
\bibitem[1998]{Letal98}
   Lefloch, B., Castets, A., Cernicharo, J., et al. 1998, \aap, 334, 269
\bibitem[1983]{Letal83}
   L\'eger, A., Gauthier, S., D\'efourneau, D., Rouan, D. 1983, \aap, 117, 164
\bibitem[1999]{Malfait}
   Malfait, K., Waelkens, C, Bouwman, J., De Koter, A., Waters, L.B.F.M.
1999, \aap, in press
\bibitem[1998]{MJC98}
   Meyer, D.M., Jura, M., Cardelli, J.A. 1998, \apj, 493, 222
\bibitem[1999]{Metal99}
   Molinari, S., et al. 1999, in preparation
\bibitem[1994]{Metal94}
   Moore, M.H., Ferrante, R.F., Hudson, R.L., et al. 1994, \apjl, 428, 81
\bibitem[1998]{NCGM98}
   Noriega-Crespo, A., Garnavich, P., Molinari, S. 1998, \aj, 116, 1388
\bibitem[1990]{Oetal90}
   Omont, A., Modeley, S.H., Forveille, T., et al. 1990, \apjl, 355, L27
\bibitem[1990]{SB90}
   Smith, M.D., Brand, P.W.J.L. 1990, \mnras, 245, 108
\bibitem[1994]{Setal94}
   Smith, R.G., Robinson, G., Hyland, A.R., et al. 1994, \mnras, 271, 481
\bibitem[1987]{SB87}
   Solf, J., B\"ohm, K.H. 1987, \aj, 93, 1172
\bibitem[1998]{WW98}
   Waters, L.B.F.M., Waelkens, C. 1998, \araa, 36, 233
\bibitem[1967]{W67}
   Wickramasinghe, N.C. 1967, ``Interstellar Grains'', Chapman \& Hall eds., London
   
   
\end{thebibliography}
\end{document}